\PassOptionsToPackage{prologue,dvipsnames}{xcolor}
\documentclass[sigconf]{aamas} 

\usepackage{balance} 
\usepackage{tikz}
\usetikzlibrary{bayesnet}
\usepackage[makeroom]{cancel}
\usepackage{cuted}
\usepackage{acro}
\usepackage[dvipsnames]{xcolor}
\usepackage{ifthen}
\usepackage{titlesec}

\makeatletter
\gdef\@copyrightpermission{
  \begin{minipage}{0.2\columnwidth}
   \href{https://creativecommons.org/licenses/by/4.0/}{\includegraphics[width=0.90\textwidth]{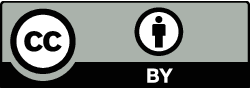}}
  \end{minipage}\hfill
  \begin{minipage}{0.8\columnwidth}
   \href{https://creativecommons.org/licenses/by/4.0/}{This work is licensed under a Creative Commons Attribution International 4.0 License.}
  \end{minipage}
  \vspace{5pt}
}
\makeatother

\setcopyright{ifaamas}
\acmConference[AAMAS '25]{Proc.\@ of the 24th International Conference
on Autonomous Agents and Multiagent Systems (AAMAS 2025)}{May 19 -- 23, 2025}
{Detroit, Michigan, USA}{Y.~Vorobeychik, S.~Das, A.~Nowé  (eds.)}
\copyrightyear{2025}
\acmYear{2025}
\acmDOI{}
\acmPrice{}
\acmISBN{}

\acmSubmissionID{fp0224}

\title[factorised-MA-AIF]{
    Factorised Active Inference for Strategic Multi-Agent Interactions
}

\author{Jaime~Ruiz-Serra}
\affiliation{
    \institution{
        Centre~for~Complex~Systems,
	The~University~of~Sydney
    }
  \city{Sydney}
  \country{Australia}
}
\email{jaime.ruizserra@sydney.edu.au}

\author{Patrick~Sweeney}
\affiliation{
    \institution{
        Centre~for~Complex~Systems,
	The~University~of~Sydney
    }
  \city{Sydney}
  \country{Australia}
}
\email{pswe2854@uni.sydney.edu.au}

\author{Michael~S.~Harré}
\affiliation{
    \institution{
        Centre~for~Complex~Systems,
	The~University~of~Sydney
    }
  \city{Sydney}
  \country{Australia}
}
\email{michael.harre@sydney.edu.au}

\begin{abstract}
    Understanding how individual agents make strategic decisions within collectives is important
    for advancing fields as diverse as economics, neuroscience, and multi-agent systems.
    Two complementary approaches can be integrated to this end.
    The \textit{Active Inference framework} (AIF) describes how agents employ a generative model to adapt their beliefs about and behaviour within their environment. 
    \textit{Game theory} formalises strategic interactions between agents with potentially competing objectives.
    To bridge the gap between the two, we propose a factorisation of the generative model whereby each agent maintains explicit, individual-level beliefs about the internal states of other agents, and uses them for strategic planning in a joint context. 
    We apply our model to iterated general-sum games with two and three players,
    and study the ensemble effects of \textit{game transitions}, where the agents' preferences (game payoffs) change over time.
    This non-stationarity, beyond that caused by reciprocal adaptation, reflects a more naturalistic environment in which agents need to adapt to changing social contexts.
    Finally, we present a dynamical analysis of key AIF quantities: the variational free energy (VFE) and the expected free energy (EFE) from numerical simulation data.
    The ensemble-level EFE allows us to characterise the basins of attraction of games with multiple Nash Equilibria under different conditions, and we find that it is not necessarily minimised at the aggregate level.
    By integrating AIF and game theory, we can gain deeper insights into how intelligent collectives emerge, learn, and optimise their actions in dynamic environments, both cooperative and non-cooperative.
\end{abstract}


\newcommand{\BibTeX}{\rm B\kern-.05em{\sc i\kern-.025em b}\kern-.08em\TeX}

\DeclareAcronym{INFG}{
    short=INFG,
    long=Iterated normal-form games
}
\DeclareAcronym{POMDP}{
    short=POMDP,
    long=Partially-Observable Markov Decision Process
}
\DeclareAcronym{AIF}{
    short=AIF,
    long=Active Inference framework
}
\DeclareAcronym{VFE}{
    short=VFE,
    long=variational free energy
}
\DeclareAcronym{EFE}{
    short=EFE,
    long=expected free energy
}
\acsetup{format/first-long=\itshape}

\begin{document}


\pagestyle{fancy}
\fancyhead{}


\maketitle 



\section{Introduction}

Collective intelligence, the emergent ability of groups to solve problems more effectively than individuals, is a phenomenon observed across biological, social, and artificial systems. 
Understanding the mechanisms that drive this collective behaviour is essential for advancing fields as diverse as neuroscience, economics, and multi-agent systems. 

Individual-level preferences incentivise the behaviour that shapes collective outcomes. 
While these preferences may not always conflict, tensions between cooperation and non-cooperation lead to emergent higher-level structures. 
Game theory models incentivised social interactions with potentially competing objectives, where a utility function maps behaviour to the real numbers. 
A Nash equilibrium represents the point where agents, independently maximising their utility, have no incentive to change their strategy.

Bridging the gap between idealised game-theoretic models and the often messy realities of agents interacting in complex environments presents a persistent challenge. 
Traditional game theory often falters when agents deviate from perfect rationality~\cite{camerer2011behavioral}. 
This challenge becomes particularly salient in the face of \textit{strategic uncertainty}, where agents grapple with uncertainty about the actions and intentions of others, and \textit{equilibrium selection}, where multiple potential equilibria exist without clear mechanisms for convergence. 
\citeauthor{Shoham2007IfMultiagent}~\cite{Shoham2007IfMultiagent} drew attention to a key issue with equilibrium selection:
    ``It seems to us that sometimes there is a rush to investigate the convergence properties, motivated by the wish to anchor the central notion of game theory in some process, at the expense of motivating that process rigorously''.

The \ac{AIF}, a process theory rooted in neuroscience, can offer a compelling perspective on these challenges.
AIF provides an empirically informed account of perception, action, and learning under uncertainty that has rapidly matured in recent years~\cite{Friston2024PixelsPlanning},
but lacks a clear framework for multi-agent strategic interactions~\cite{Friston2024FederatedInference, Demekas2024AnalyticalModel}.
Game theory often assumes perfect rationality and complete information, which \ac{AIF} relaxes.
Employing game theory as a model of incentivised decision-making, and \ac{AIF} as the cognitive process underlying individual decisions allows for experiments with dynamic agent preferences while providing access to how their beliefs and precision change in response to others.
Our results shed light on the mutual influence between individual cognition and structural dynamics at the collective level.


We begin by reviewing recent work on the intersection of AIF and Bayesian agents, game theory, and multi-agent systems (\S\ref{sec:background}).
Integrating the two ends of the spectrum, we propose a factorisation of the generative model whereby an agent maintains explicit, individual-level beliefs about the internal states of other agents and uses them for strategic planning in a joint context (\S\ref{sec:model-description}).

We apply our model to iterated general-sum games with two and three players and study the ensemble effects of \textit{game transitions}, where the agents' preferences (game payoffs) and their associated equilibria change over time (\S\ref{sec:results}).
We present a dynamical analysis of two key AIF quantities: the \ac{VFE} (\S\ref{sec:results-vfe}) and the \ac{EFE} (\S\ref{sec:results-efe}) from numerical simulation data.
The ensemble-level EFE allows us to characterise the basins of attraction of games with multiple Nash Equilibria (such as the Stag Hunt) under different conditions, and we find that it is not necessarily minimised at the aggregate level \cite{Hyland2024FreeEnergyEquilibria}.

\section{Background}\label{sec:background}

\subsection{Iterated normal-form games}\label{sec:infg}

\ac{INFG}~\cite{Hannan1957ApproximationBayes} provide a structured framework to study strategic interactions between agents, allowing for the analysis of decision-making processes over repeated encounters.
\ac{INFG} extend the basic framework of normal-form games, where players (agents) simultaneously choose strategies (actions), and their payoffs depend on the combination of all chosen strategies. 
In an iterated setting, this process repeats over multiple rounds, allowing agents to observe outcomes and potentially adapt their strategies over time. 
\ac{INFG} are defined by a set of agents, a set of allowable actions for each agent,~$\mathcal{U}_i$, a game payoff function, ~$\texttt{g}$, mapping every joint outcome (actions of all players involved) to a real number (payoff value for a given outcome)---thus encoding (often as a matrix) the incentives or preferences of the agents---, and possibly the total number of rounds (time steps) the game is played for.
We use the term `ego' to refer to any arbitrary agent from whose perspective we are describing the game, and `alter' to refer to any other agent participating in the interaction.


The simplest games are (symmetric) two-player, two-action ($2\times 2$) games\footnote{
Two-action settings (e.g., cooperate/defect) are standard in game theory due to their simplicity, analytical tractability, and clarity of incentives. 
These settings focus on fundamental dynamics and are widely applicable to real-world scenarios abstracted into binary decisions.
}.
Here actions are $u \in \mathcal{U} = \{0, 1\} \equiv \{\texttt{c}, \texttt{d}\}$ (`cooperate' and `defect') for each agent.
Ego's payoffs for each of the four possible outcomes in these games are commonly referred to as
\textit{reward} ($R$) when both agents cooperate, 
\textit{temptation} ($T$) when ego defects and alter cooperates, 
\textit{sucker} ($S$) when ego cooperates but alter defects, 
and \textit{penalty} ($P$) when both defect.
From ego's point of view, her payoffs are:
\begin{equation}
    \texttt{g} =
    \begin{bmatrix}
      R & S \\ 
      T & P 
    \end{bmatrix}
\end{equation}

Canonical games can be determined by the relative ordering of these payoff values~\cite{Harre2018MultiagentEconomics}, for example
the well-known \textit{Prisoner's Dilemma} (\texttt{PD}) has $T>R>P>S$, 
the \textit{Chicken} game (\texttt{Ch}) has $T>R>S>P$, 
and the \textit{Stag Hunt} (\texttt{SH}) has $R>T>P>S$. 
By setting each as an integer between 1 and 4 as in~\cite{Demekas2024AnalyticalModel}, we have:
\begin{equation}\label{eq:2x2-payoffs}
    \texttt{PD} =
    \begin{bmatrix}
      3 & 1 \\ 
      4 & 2 
    \end{bmatrix},
    \;
    \texttt{Ch} =
    \begin{bmatrix}
      2 & 3 \\ 
      4 & 1 
    \end{bmatrix},
    \;
    \texttt{SH} =
    \begin{bmatrix}
      4 & 1 \\ 
      3 & 2 
    \end{bmatrix},
\end{equation}
from which we can obtain the payoff function, e.g. 
$$\texttt{g}_{\texttt{Ch}}(\texttt{d}, \texttt{c}) = T_{\texttt{Ch}} = 4.$$

In the single-shot version of a normal-form game, agents typically maximise their payoff for that round alone. 
However, in iterated games, agents must consider long-term outcomes~\cite{fudenberg1991repeated}.
This opens up new possibilities for strategic behaviour, including learning (agents can learn from previous interactions adjusting their strategy to improve future outcomes),
reciprocity (agents might cooperate if they believe others will reciprocate in future rounds, balancing short-term losses for long-term gains), 
and reputation and trust (the interaction history can influence future decisions, where agents may choose to trust or punish based on previous behaviour)~\cite{axelrod1981evolution}.


\subsection{Bayesian learning in games}

Bayesian learning in strategic games builds on Savage's foundational work in Bayesian decision theory, which originally addressed games against nature~\cite{savage1954thefoundationsof}. In multi-agent settings, outcomes depend on the strategic interplay of agents, where each agent’s actions influence and respond to those of others. This interdependence creates a dynamic, non-stationary environment, requiring agents to adapt continually as strategies coevolve~\cite{hernandez2017survey, albrecht2018autonomous, li2022role}. 

The simulation literature focuses on how agents can learn equilibria through repeated interactions. The choice of priors significantly influences which equilibria agents achieve~\cite{fudenberg1993self, nyarko1994bayesian, dekel2004learning}. Under certain conditions, rational learning may converge asymptotically to a Nash equilibrium if agents’ priors contain a ‘grain of truth’~\cite{kalai1993rational, jordan1991bayesian, nachbar2005beliefs}. A fundamental model in this area is fictitious play, where agents estimate opponents’ strategies by averaging past actions and choosing their best response~\cite{brown1951iterative, robinson1951iterative}. This framework can be interpreted as sequential Bayesian inference, where each agent assumes opponents follow an unknown, independent, and stationary strategy~\cite{young2004strategic}. Extensions of fictitious play introduce stochastic action selection~\cite{fudenberg1995consistency, fudenberg1999conditional, mckelvey1995quantal}, exponential forgetting~\cite{fudenberg1998theory}, non-stationary strategies~\cite{shamma2005dynamic}, and variational inference~\cite{Rezek2008SimilaritiesInference}. 

In AI, the multi-agent systems literature explores Bayesian methods for coordination and learning, though they have not yet gained the same traction as in single-agent reinforcement learning~\cite{ghavamzadeh2015bayesian}. A common approach is to adapt single-agent algorithms, such as the Bayes-Adaptive Markov Decision Process~\cite{duff2002optimal} and its extensions~\cite{ross2007bayes, chalkiadakis2003coordination, guez2012efficient, rigter2021risk}. These algorithms, however, often struggle in multi-agent environments due to the non-stationarity introduced by agents’ coevolving strategies, making it challenging for any single agent to converge to an optimal policy.

To address these challenges, type-based reasoning has emerged as a prominent approach for explicitly modelling other agents. In this approach, agents model others’ behaviours as mappings from interaction histories to action probabilities~\cite{harsanyi1967games, albrecht2016belief}, allowing them to anticipate and respond to a wide range of strategies. This method addresses the heterogeneity of multi-agent systems by classifying agents according to their learning capabilities and information structures~\cite{li2022role}. By starting with a prior over these types, agents systematically update their beliefs based on observed actions, refining their predictions and strategies as they gather more information about their opponents’ behaviors~\cite{carmel1999exploration, southey2012bayes, hoang2013general, albrecht2015game, wu2021too}. 

Recursive reasoning builds on type-based methods by incorporating not only an agent's beliefs about others but also their beliefs about the beliefs of others, forming a hierarchical structure. This approach underlies models like the Interactive Partially Observable Markov Decision Process \cite{gmytrasiewicz2000rational, gmytrasiewicz2005framework, gmytrasiewicz2004interactive}, where agents maintain and update beliefs about others’ beliefs and strategies across multiple levels. By modelling nested beliefs, agents better anticipate others’ actions, laying the foundation for Theory of Mind models that support more sophisticated and adaptive interactions~\cite{baker2011bayesian, ng2012bayes, panella2017interactive, doshi2020recursively, wu2021too}.

Another line of research extends graphical models to multi-agent contexts, leveraging conditional independencies for efficient representation and inference. Multi-agent influence diagrams and graphical games capture dependencies among agents, enabling efficient computation by focusing on local interactions~\cite{koller2001structured, koller2003multi, kearns2013graphical}. Similarly, action-graph games and expected utility networks optimise inference by structuring interactions around shared actions within agent subsets, which is particularly advantageous in sparsely coupled games~\cite{jiang2010bayesian, jiang2011action, la2013expected}.

Finally, the intersection of Bayesian learning and bounded rationality frames decision-making as constrained optimisation under uncertainty. Product Distribution Theory applies the Maximum Entropy principle \cite{jaynes2003probability} to derive equilibria where agents balance utility maximisation with computational costs~\cite{Wolpert2006InformationTheory}. \citeauthor{grunwald2004game} demonstrate that maximising entropy and minimising worst-case expected loss are dual problems, structured as zero-sum games between a decision-maker and nature~\cite{grunwald2004game}. Thermodynamic Decision Theory expands these principles, integrating utility (energy) and information-processing costs (entropy) within a variational framework where agents minimise free energy. This approach naturally extends to variational Bayesian inference, enabling efficient, approximate posterior updates under bounded rationality constraints~\cite{ortega2013thermodynamics}. This framework generalises to risk-sensitive control \cite{kappen2012optimal, levine2018reinforcement, fellows2019virel, o2021variational, o2020making} and adversarial contexts ~\cite{ortega2011information, ortega2014adversarial}, illustrating its versatility across diverse decision-making scenarios.

\subsection{Game theory and Active Inference}

This section outlines how \ac{AIF} and game theory have been combined to model strategic decision-making in social interactions. 
\citeauthor{Yoshida2008GameTheory}~\cite{Yoshida2008GameTheory} examine how individuals infer the intentions of others in the spatial Stag Hunt game, highlighting that people engage in recursive thinking about others' beliefs. This approach connects strategic thinking in game theory with Bayesian inference and bounded rationality in cognitive psychology, providing insights into how people make decisions in uncertain social environments.

\citeauthor{Moutoussis2014FormalModel}~\cite{Moutoussis2014FormalModel} develop a formal model of interpersonal inference based on active inference principles, where agents infer their partner’s likely type (cooperative or defecting) by observing past actions and updating their beliefs. This demonstrates how the computational models used to describe individual decision-making can be extended to the complexities of social interaction, where understanding others' mental states is crucial.

\citeauthor{Demekas2024AnalyticalModel}~\cite{Demekas2024AnalyticalModel} build on this by showing how AIF agents can learn effective strategies in the iterated Prisoner's Dilemma by continuously updating beliefs about transition probabilities between game states. Their generative model tracks how learning rates influence strategy development, offering analytical clarity through belief updates. \citeauthor{Hyland2024FreeEnergyEquilibria}~\cite{Hyland2024FreeEnergyEquilibria} introduce the `Free-Energy Equilibria' framework, extending the EFE to strategic contexts by conditioning predictions on the joint policies of agents. This framework merges Nash equilibria with bounded rationality, proposing that cooperation could emerge as agents align actions through joint free-energy minimization.

\citeauthor{Fields2024NashEquilibria}~\cite{Fields2024NashEquilibria} further explore how physical interactions can be framed as games using the Free Energy Principle, drawing attention to the undecidability of achieving Nash equilibria in classical and quantum contexts. This complexity helps explain why real-world systems often fail to converge to stable outcomes.

Grounding strategic decision-making in AIF provides a more realistic model for social interactions, moving beyond the assumptions of perfect rationality and complete information in traditional game theory, with a foundation in neuroscience.

\section{Model description}\label{sec:model-description}

In \ac{INFG}, $N$ agents interact by selecting an action each time step with the goal of maximising their payoffs as determined by the game payoff function, $\texttt{g}$.  
The agents observe the actions taken by each of the agents (including themselves) in the preceding step in order to decide how to act in the current step. 

In our model\footnote{
Code available at \href{https://github.com/RuizSerra/factorised-MA-AIF}{GitHub/RuizSerra/factorised-MA-AIF}
}, these observations are perceived via different \textit{modalities}, $m$, one for each agent.
For example, for $N = 3$ agents $m\in\{i, j, k\}$, $|\mathcal{U}|=2$ actions, and taking an egocentric perspective,
each observation is $\mathbf{o} = (o_i, o_j, o_k) \in \{\texttt{c}, \texttt{d}\}^N$, with the first modality being \textit{ego}'s action $o_{i,t} = u_{i,(t-1)}$, and subsequent modalities $o_{j,t} = u_{j, (t-1)}$ each pertaining to an \textit{alter} ($j, k$).
Further details are provided in \S\ref{sec:factorised-model}.
In the following, we omit time and agent indices where implied by context. 

To act effectively in response to her counterparts, ego must take into account each of her opponents' propensity for playing each action at a given time (e.g. for $j$ to play `cooperate', or $p(u_j = \texttt{c})$).
This propensity is driven by the opponent's `internal world', $\psi_j$, which is not observable to ego~\cite{lombard2023causal}.

An appropriate way to model $\psi_j$ is as a hidden state using a \ac{POMDP}, where, each time step, agents infer the current hidden state $s \in \mathcal{S}$ based on an observation $o \in \mathcal{O}$, and can influence it through their actions $u \in \mathcal{U}$ to maximise their payoff.
\ac{AIF} distinguishes between the external (ontological) \textit{generative process}, which represents the actual dynamics of the environment, and the internal (epistemic) \textit{generative model} of each agent, which encodes its beliefs about those dynamics, and as such is a good match for the \ac{POMDP} formalism~\cite{DaCosta2020ActiveInference, Smith2022StepstepTutorial}.

\subsection{Generative Model}\label{sec:generative-model}

An agent's generative model consists of a joint distribution over hidden states, observations, policies (action sequences), and model parameters. 
The short timescale dynamics are encoded in
$$
p(s_{0:t}, o_{0:t} | u_{0:t-1}) = 
    p(s_0) 
    \prod_{\tau=1}^t 
        p(o_\tau|s_\tau) 
        p(s_\tau|s_{\tau-1}, u_{\tau-1}),
$$
including the agent's prior beliefs about the initial state of the world $p(s_0)$ encoded in $\boldsymbol{\mathsf{D}}$; the transition model $p(s_{t+1}|s_t, u_t)$, encoded in $\boldsymbol{\mathsf{B}}$; and the observation likelihood $p(o|s)$, encoded in $\boldsymbol{\mathsf{A}}$.
For discrete POMDPs, the distributions can be obtained from their encoding as a categorical distribution, e.g. $p(o|s) = \operatorname{Cat}(\boldsymbol{\mathsf{A}})$ where $\boldsymbol{\mathsf{A}}$ is a $|\mathcal{O}|\times |\mathcal{S}|$ matrix, $\boldsymbol{\mathsf{B}}$ is a $|\mathcal{S}|\times |\mathcal{S}| \times |\mathcal{U}|$ tensor, and $\boldsymbol{\mathsf{D}}$ a vector in the simplex $\Delta_\mathcal{S}$.

\begin{figure}
    \centering
    \includegraphics[width=1\linewidth]{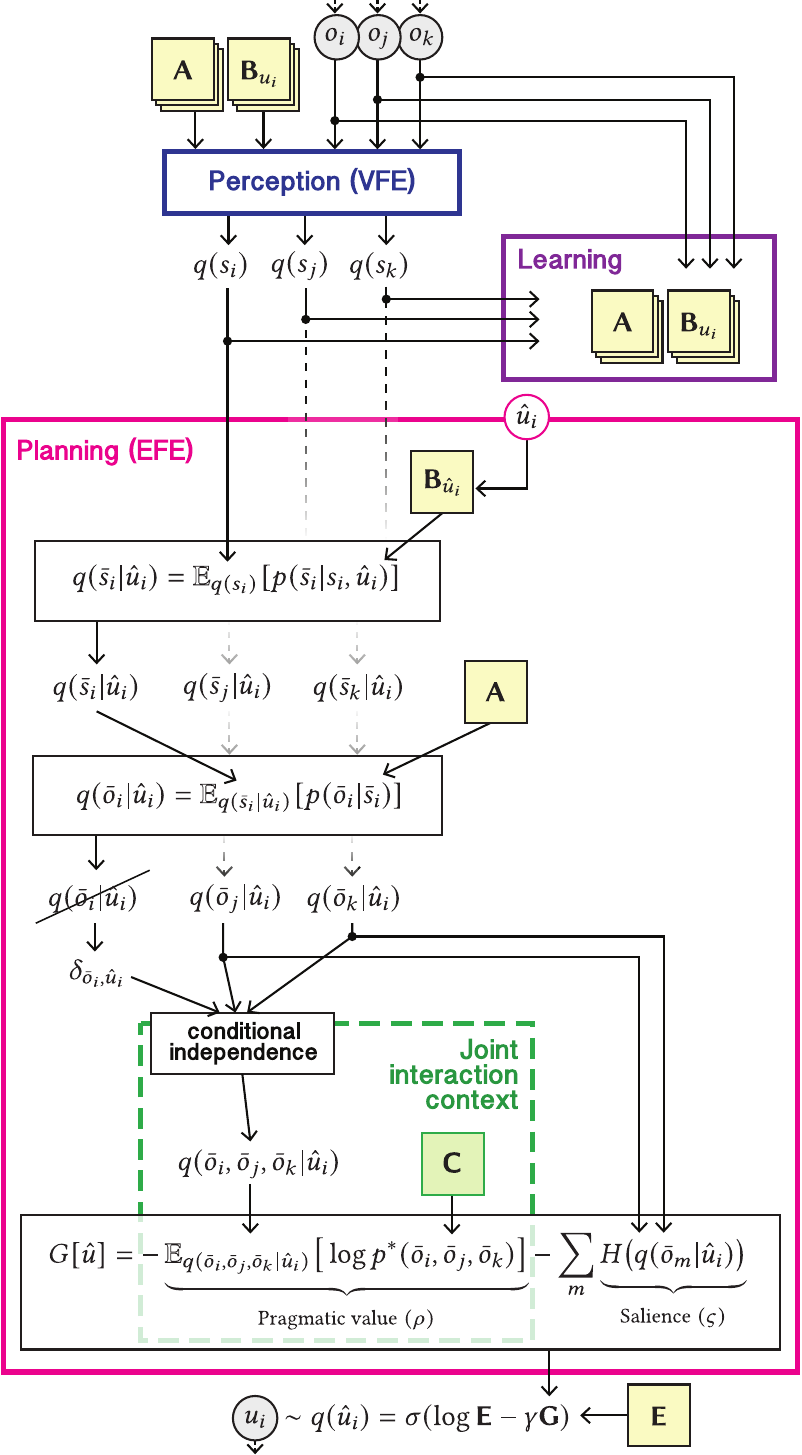}
    \caption{
        The perception-action loop;
        ego's `internal world', $\psi_i$. 
        The agent observes the actions of all agents, from which she updates her beliefs $q(s)$ to minimise VFE.
        These beliefs are used to plan her next action by minimising EFE.
        }
    \label{fig:computational-graph}
\end{figure}

In our application to strategic interactions, ego infers the hidden state of each agent~\cite{Ruiz-Serra2023InverseReinforcement, baker2011bayesian} in a corresponding \textit{factor}, $n \in \{i, j, k\}$, of the generative model.
Ego must make vast simplifications in modelling each alter's true and complex internal world $\psi_j$---which encompasses all the components in alter's generative model, encoding his \textit{beliefs} ($\boldsymbol{\mathsf{A}}, \boldsymbol{\mathsf{B}}, \boldsymbol{\mathsf{D}}, q$; see \S\ref{sec:variational-inference}), \textit{preferences} ($\boldsymbol{\mathsf{C}}$; see \S\ref{sec:learning-preferences}), and \textit{constraints} (such as habits $\boldsymbol{\mathsf{E}}$, or level of rationality $\beta_1$; see \S\ref{sec:action-selection})~\cite{Gintis2006FoundationsBehavior}, all depicted in Figure~\ref{fig:computational-graph}.
Our approach is to model this internal world (external to ego) as a categorical distribution over different types, with parameters $\mathbf{s}$, placing the opponent type on the simplex $\Delta_{\mathcal{S}}$.
In the simplest such model, which we adopt here, these types may be `cooperator' or `defector', and the opponent could be anywhere between the two (i.e., we model the opponent's propensity for playing each action)\footnote{
The support of this distribution need not be limited to the number of actions. For example, an agent could consider four possible types for policy length 2, or more abstract types such as `Tit for Tat'~\cite{axelrod1981evolution}.
}.

Ego further assumes that her opponents play the actions they mean to play, ruling out the possibility of `trembling hand' imperfections. 
This simplifies our model such that there is no `ambiguity' in the environment, so the likelihood model for each factor $\boldsymbol{\mathsf{A}}_n$ is an identity matrix, 
or equivalently that the observation likelihood is a Kronecker delta distribution, $p(o_m|s_n) = \delta_{o_m, s_n}$, $\forall\,m=n\in\{i, j, k\}$.

\subsection{Variational inference}\label{sec:variational-inference}

Every time step, having observed the actions of each agent, $o_n = u_n$, ego has to infer the underlying $s_n \equiv p(u_n)$, for each factor.
This entails updating her posterior beliefs $p_i(s_j | o_j)$ about each hidden state factor through Bayesian inversion. 
The true posterior may be intractable, so it is approximated by a variational posterior $q_i(s_j)$.
This is achieved by minimising the \textit{Variational Free Energy} (VFE), which is expressed as (having omitted the agent $i$ and factor $j$ indices for brevity):
\begin{subequations}
    \begin{align}
        F[q, o] &= \label{eq:vfe-energy-entropy}
        \underbrace{\mathbb{E}_{q(s)}[-\log p(o, s)]}_{\text{energy}} 
        -\underbrace{H[q(s)]}_{\text{entropy}} 
        \\
    &= 
        \underbrace{\operatorname{D}_{\textsl{\textsc{kl}}}\big[q(s) \,\big\|\, p(s|o)\big]}_{\text{divergence}}
        -\underbrace{\log p(o)}_{\text{evidence}}
    \ge
        -\underbrace{\log p(o)}_{\text{evidence}}
    ,
    \end{align}
\end{subequations}
with $p(s)$ the prior over hidden states, and $q(s)$ the variational posterior, or the agent's beliefs about hidden states.
We model beliefs via a $\operatorname{Dirichlet}(\boldsymbol{\theta})$ distribution with variational parameters~$\boldsymbol{\theta}$.
We approximate the \ac{VFE} using the following Monte Carlo sampling procedure:
let $q_l\sim\operatorname{Dirichlet}(\boldsymbol{\theta})$ be the $l$-th sample from the Dirichlet distribution.
With $L$ samples $\{q_l\}_{l=1}^L$, we can write an unbiased estimate of the \ac{VFE}\footnote{
The chosen form of the VFE for optimisation is derived from (\ref{eq:vfe-energy-entropy}) by replacing $p(o,s) = p(o|s)p(s)$ and subsuming all terms into a single expectation operator in (\ref{eq:vfe-expectation}).
} as
\begin{subequations}
    \begin{align}
        \label{eq:vfe-expectation}
        F[q, o] &= -\mathbb{E}_{q(s)}\big[
    \log p(o|s)
    + \log p(s)
    - \log q(s)
\big]\\
        & \approx  
        \hat{F}\big[\{q_l\}_{l=1}^L, o\big] = 
        \frac{1}{L} \sum_{l=1}^{L} F[q_l, o]
    \end{align}
\end{subequations}
and optimise the variational parameters $\boldsymbol{\theta}$ through stochastic gradient descent on $\hat F$. 
Having found $\boldsymbol{\theta}^*$ upon completing the procedure, 
we can recover the inferred posterior $q(s)$ as the expected value of $\operatorname{Dirichlet}(\boldsymbol{\theta}^*)$, which serves as the optimal point estimate under a quadratic loss function \cite{bernardo2009bayesian}.

Since variational inference occurs every time step, the prior $p(s)$ is obtained from the previously inferred state and the transition model $\boldsymbol{\mathsf{B}}_u$, i.e. 
\begin{equation}\label{eq:state-transition}
    p(s_t) \approx q(s_t|u_{t-1}) = 
    \sum_{s_{t-1}} p(s_t|s_{t-1}, u_{t-1}) q(s_{t-1})
\end{equation}


\subsubsection{Factorised model}\label{sec:factorised-model}


Generative models in previous AIF-adjacent applications to game theory \cite{Moutoussis2014FormalModel, Demekas2024AnalyticalModel, Freire2019ModelingTheory} assume the state space is joint across all agents in the game (i.e. \{\texttt{cc}, \texttt{cd}, ... \texttt{dd}\} for their two agents).
Is a mean-field factorisation of the variational posterior,  $q(s_i)q(s_j)q(s_k)$, adequate in the game-theoretic context, or should a joint distribution, $q(s_i, s_j, s_k)$, be assumed ~\cite{Rezek2008SimilaritiesInference}?
That is, can the hidden states be considered independent of each other?
Recall that, when our agents perform inference, they approximate the posterior $p(s|o) \approx q(s)$. 
In the case of repeated normal-form games, the hidden state $s_j$ is (the parameterisation of) the posterior distribution over actions (policy) $q(u_j)$ of opponent $j$, and the observation $o_j$ is the action $u_j$ taken by this opponent.
So inferring a distribution over a single opponent's hidden state involves finding a $q(s_j) \approx p(s_j | o_i, o_j, o_k)$.

The \textit{Markov Blanket} (MB) of a node or set of nodes $\mathcal{J}$ is defined as the set of $\mathcal{J}$'s parents, $\mathcal{J}$'s children, and any other parents of $\mathcal{J}$'s children \cite{Pearl2014ProbabilisticReasoning}.
Under this definition, if we let $\mathcal{J}$ be the nodes `inside' agent $j$, including $s_j = q(u_j)$, the MB consists of $\{o_i, o_j, o_k, u_j\}$, which `shield' the internal states of $j$ from the external world. 
In \ac{INFG}, taking dynamics into account, this MB set is actually $\{u_{i(t-1)}, u_{j(t-1)}, u_{k(t-1)}, u_{jt}\}$.
Thus, we can say that 
\begin{equation*}
    s_j \perp \{s_i, s_k\} \;\big|\; \{u_{i(t-1)}, u_{j(t-1)}, u_{k(t-1)}, u_{jt}\}, 
\end{equation*}
i.e., $s_j$ is \textit{conditionally independent} of $s_i$ and $s_k$ (and any other node outside $j$) \textit{given} $\{u_{i(t-1)}, u_{j(t-1)}, u_{k(t-1)}, u_{jt}\}$. 
Furthermore, at time $t$ when $i$ infers $s_j$, the action $u_{jt}$ has not happened yet, so that node does not exist in the Dynamic Bayesian Network. 
Therefore, $q(s_j) \approx p(s_j | o_i, o_j, o_k)$, i.e. given the MBs of agents, their internal states are conditionally independent.

Accordingly, in our model, each agent $i$ infers the hidden state for every factor $n\in \{i, j, k\}$ individually, and thus retains a collection of parameters 
$\boldsymbol{\theta} \in \mathbb{R}_+^{N \times |\mathcal{U}|}$,
with $N$ the number of factors (agents being tracked) and $|\mathcal{U}|$ the number of actions.




\subsection{Preferences and planning}\label{sec:learning-preferences}

AIF extends Bayesian learning by integrating action and decision-making, allowing agents to actively reduce uncertainty and achieve goal-directed behaviour in dynamic environments.
Agents plan and select actions that both gather information and satisfy their preferences over observations $p^*(o)$, encoded in $\boldsymbol{\mathsf{C}}$ (such that $p^*(o) = \operatorname{Cat}(\boldsymbol{\mathsf{C}})$ in discrete settings). 
This requires counterfactual thinking: `what would I be likely to observe if I were to do $u_i$?'.
The generative model employed in inference can be used for planning by predicting future states (via the transition model $\boldsymbol{\mathsf{B}}_{\hat u}$) and observations (via the likelihood $\boldsymbol{\mathsf{A}}$) given counterfactual actions $\hat u_i$ (distinguished from actual actions $u_i$).
In the following, we denote predictive variables with a bar, e.g. $\bar o_j$ is what $i$ might observe $j$ doing in the next time step.

Agents achieve the desired exploration-exploitation trade-off by selecting actions that minimise an \textit{Expected Free Energy} (EFE), 
comprising \textit{salience} and \textit{pragmatic value} terms.
In what follows, we describe these terms in more detail and adapt them to INFG.

\subsubsection{Salience}

Otherwise known as `epistemic value', salience ($\varsigma$) captures the information gain about hidden states---or how an action is anticipated to change one's beliefs---with greater changes in beliefs holding higher epistemic value~\cite{Parr2022ActiveInference}.
It can be decomposed into a difference between two entropic terms: 
\begin{align}
    \varsigma[\hat u] &= \mathbb{E}_{q(\bar o| \hat u)}\big[
            \operatorname{D}_{\textsl{\textsc{kl}}}\big[
                q(\bar s| \hat u, \bar o) 
                \,\big|\big|\,
                q(\bar s | \hat u)
            \big]
        \big]\\
    &=
    H\big(q(\bar o| \hat u)\big)
            - 
            \mathbb{E}_{q(\bar s| \hat u)}\big[
                H\big(p(\bar o| \bar s)\big)
            \big]
\end{align}
Since our INFG environment is unambiguous (cf. \S\ref{sec:generative-model}), the second term (ambiguity) is zero.
There is, however, information to be gained still from acting to maximise the first term, leading to exploratory behaviour.
This term captures the uncertainty about the next observation in the event that $\hat u_i$ is the action taken.
Actions whose outcomes we are most uncertain about are preferred, as we stand to gain the most information from them.
Salience is additive over factors:
\begin{equation}
    \varsigma[\hat u_i] =
    \sum_{m}
    H\big(q(\bar o_m| \hat u_i)\big)
\end{equation}

\subsubsection{Pragmatic value} 

By choosing actions that maximise pragmatic value ($\rho$), agents actively pursue their preferences, closing the gap between predicted and preferred observations. 
In normal-form games, preferences are determined by the game payoffs $\texttt{g}$, and we can convert directly from one to the other with
\begin{equation}\label{eq:preferences-payoffs}
    p^*(o_i, o_j, o_k) = \sigma\big(\texttt{g}(o_i, o_j, o_k)\big)
\end{equation}
$\forall (o_i, o_j, o_k) \in \mathcal{U}^N$, where $\sigma$ is the \textit{softmax} function\footnote{
The \textit{softmax} operation here is not strictly necessary; we could just as well go with $p^*(o_i, o_j, o_k) = \exp\big( \texttt{g}(o_i, o_j, o_k) \big)$ under an energy function interpretation. 
However, by ensuring $p^*$ is a probability distribution, we ensure the EFE values are in the positive range (i.e. in Nats) \cite{Lecun2006TutorialEnergybased, Gottwald2020TwoKinds}.
}~\cite{Demekas2024AnalyticalModel, Ortega2009ConversionUtility}.

This implies a \textit{joint interaction context}, since the preferences for each observation modality come from a joint distribution and are generally not independent.
This highlights the core principle of game theory, namely the interdependence of agents' preferences and actions.
Here the preferences for each observation modality are derived from a joint distribution, meaning that agents' outcomes are interconnected, and their strategies cannot be considered in isolation. 
The essence of game theory lies in analyzing how these dependencies shape decision-making and the resulting equilibria in strategic interactions.
Agents track the `mental states' of other agents \textit{individually} (in $q(s_j)$), but they must learn how they interplay in a given joint context defined by $\texttt{g}$ (e.g. when \textit{two} agents play a \textit{Prisoner's Dilemma}, or when \textit{three} agents play \textit{Chicken}).
Unlike in classical game theory, however, our agents do not know the payoff function (preferences) of their opponents.

The predicted observations, as a posterior predictive distribution $q(\bar o| \hat u)$ for each modality, need to be merged into a single $q(\bar o_i, \bar o_j, \bar o_k | \hat u_i)$ to be able to compare to the preferences.
The conditional independence between the internal states of the agents, determined by their respective MBs (\S\ref{sec:factorised-model}), allows us to define this joint posterior as the product~\cite{Wolpert2006InformationTheory}
\begin{equation}
    q(\bar o_i, \bar o_j, \bar o_k | \hat u_i) 
    = \prod_{m \in \{i, j, k\}} 
        q(\bar o_m | \hat u_i) 
\end{equation}
where each factor's posterior predictive observation is obtained from the (factor's) likelihood model $\boldsymbol{\mathsf{A}}_j$ as\footnote{
$q(\bar o_j | \hat u_i) = \sum_{\bar s_j} q(\bar o_j, \bar s_j | \hat u_i) = \sum_{\bar s_j} p(\bar o_j | \bar s_j) q(\bar s_j | \hat u_i) = \mathbb{E}_{q(\bar s_j | \hat u)}[p(o_j | \bar s_j)]$
} 
\begin{equation}
    q(\bar o_j | \hat u_i) 
    = \mathbb{E}_{q(\bar s_j | \hat u)}[p(\bar o_j | \bar s_j)].
\end{equation}
The pragmatic value is thus defined as the negative cross-entropy between the posterior predictive observation and the preference distributions, which agents aim to maximise (cf. log-loss minimisation):
\begin{equation}
    \rho [\hat u_i] = \mathbb{E}_{q(\bar o_i, \bar o_j, \bar o_k | \hat u_i)}\big[\log p^*(\bar o_i, \bar o_j, \bar o_k)\big].
\end{equation}

Furthermore, for the ego's own factor, $\bar o_i = \hat u_i$ is guaranteed with full certainty in the counterfactual where $\hat u_i$ is played (i.e. where $u_i$ would in actuality be $\hat u_i$). Accordingly, we can set $q(\bar o_i | \hat u_i) = \delta_{\bar o_i, \hat u_i}$, the Kronecker delta distribution\footnote{or, equivalently, $\mathbf{1}(\hat u_i)$, the one-hot encoding of the action under consideration}. 
This makes the pragmatic value
\begin{equation}
    \rho[\hat u] \equiv
    \mathbb{E}_{q(\bar o_j)q(\bar o_k)}[\log p^*(\bar o_i, \bar o_j, \bar o_k)|\bar o_i = \hat u_i],
\end{equation}
i.e., equivalent to the game-theoretic \textit{expected utility}, under the interpretation that $\log p^*$ is (proportional to) the utility function (i.e., the game payoffs, as we did in Eq.~\ref{eq:preferences-payoffs}). 
Finally, we have the EFE,
\begin{equation}
    G[\hat u_i] = - \rho[\hat u_i] - \varsigma[\hat u_i] ,
\end{equation}
which the agents minimise through their actions, effectively maximising salience ($\varsigma$) and pragmatic value ($\rho$).
For the zero-ambiguity case under consideration, the EFE reduces to (in shorthand)
\begin{equation}
    G[\hat u_i] = - \mathbb{E}_{q_{-i}}[\log p^*|\hat u_i] - \sum_m H\big(q_m\big),
\end{equation}
highlighting the relationship with previous information-theoretic treatments of bounded rationality in game theory~\cite{Wolpert2006InformationTheory, Ortega2015InformationTheoreticBounded}.

\subsection{Action selection}\label{sec:action-selection}

The selection of actions is driven by the precision-modulated EFE of each possible action $G[\hat u_i]$ (or $\boldsymbol{\mathsf{G}}$ in vector notation), and the agent's habits $\boldsymbol{\mathsf{E}}$ (uniform):
\begin{equation}
    u_i \sim q(\hat u_i) = \sigma\big(\log \boldsymbol{\mathsf{E}} - \gamma \boldsymbol{\mathsf{G}} \big),
\end{equation}
where $\gamma$ is a precision parameter updated each time step,
\begin{equation}
    \gamma 
    =
    \frac{
        \beta_1
    }{
        \beta_0 - \langle \boldsymbol{\mathsf{G}} \rangle
    },
\end{equation}
with fixed hyperparameters $(\beta_0, \beta_1)$.
$\beta_0$ (shape) represents a baseline level of uncertainty or noise, and 
$\beta_1$ (rate) reflects how strongly the agent's precision (or confidence) is influenced by environmental feedback. 
Intuitively, $\beta_0$ controls the threshold at which the agent starts doubting its action selections, while $\beta_1$ modulates the rate of precision updating based on the difference between expected and observed outcomes. 
Higher values of $\beta_1$ correspond to greater sensitivity to discrepancies, making the agent more `rational' and noise-averse in refining its action policy \cite{Friston2015ActiveInferenceb, friston2016active}.
External feedback is accounted for in $\langle \boldsymbol{\mathsf{G}} \rangle = \mathbb{E}_{q(\hat u_i)}\big[G[\hat u_i]\big]$, the expected EFE under the current action probabilities for this agent.

\subsection{Learning}\label{sec:learning}

Learning in AIF entails updating model parameters, and occurs at a slower rate than inference.
In our model, agents update the transition model $\boldsymbol{\mathsf{B}}$ (initially uniform) every $T_L$ steps, where $T_L$ is an integer sampled uniformly from $18 \le T_L \le 30$ each time learning occurs.
This random offset ensures agents do not learn in lockstep, which might cause artifacts in the dynamics.
The parameters are updated based on past transitions, $h_{t:t+T_L} = (\mathbf{s}_t, u_{i,t}, \mathbf{s}_{t+1}, u_{i,t+1}, ..., \mathbf{s}_{t+T_L})$, via
\begin{equation}
    \boldsymbol{\mathsf{B}}'_{u,n} = \boldsymbol{\mathsf{B}}_{u,n} +
    \sum_{\tau=t}^{t+T_L-1} 
        \alpha_l\, \delta_{u, u_{i,\tau}} (\mathbf{s}_{n,\tau+1} \otimes \mathbf{s}_{n,\tau})
\end{equation}
for each factor $n$, with learning rate $\alpha_l = 1$, and where $\otimes$ is the outer product and $\mathbf{s}_{n,t}$ sufficient statistics for $q(s_n)$ at time $t$. We refer the reader to \cite{DaCosta2020ActiveInference} for further details.

\subsubsection{Novelty}
With learning present, the agents can consider how their actions are likely to influence their generative model. 
In this case, an agent predicts how the transition model might change if they were to play action $\hat u_i$ (for each factor $n$):
\begin{equation}
    \boldsymbol{\mathsf{\bar B}}_{\hat u_i, n} = 
        \boldsymbol{\mathsf{B}}_{\hat u_i, n} 
        + \alpha_l\, (\mathbf{\bar s}_n \otimes \mathbf{s}_n) 
\end{equation}
Novelty is an additional term in the EFE, constituting additional epistemic value:
\begin{equation}
    \eta[\hat u_i] = \sum_n \operatorname{D}_{\textsl{\textsc{kl}}}\big[
        \boldsymbol{\mathsf{\bar B}}_{\hat u_i, n}
        \,\big|\big|\,
        \boldsymbol{\mathsf{B}}_{\hat u_i, n}
    \big]
\end{equation}
summed over factors.
Including novelty in the EFE, we have
\begin{equation}
    G[\hat u_i] = - \rho[\hat u_i] - \varsigma[\hat u_i] - \eta[\hat u_i]
\end{equation}  

\subsubsection{Bayesian model reduction}

A Bayesian model reduction procedure~\cite{DaCosta2020ActiveInference} applied to $\boldsymbol{\mathsf{B}}$ (for each factor and action) at learning time helps reduce overfitting. 
A `reduced' model $\boldsymbol{\mathsf{\tilde B}}_{u, n} = \sigma(\alpha_r^{-1}\boldsymbol{\mathsf{B}}_{u, n} )$ is proposed (with reduction rate $\alpha_r = 1.25$) and their evidence difference is computed as 
\begin{equation}
    \log \tilde p(o_n) - \log p(o_n) 
    = 
    \log \mathbb{E}_{\boldsymbol{\mathsf{B}}'_{u, n}} \bigg[
        \frac{
            \boldsymbol{\mathsf{\tilde B}}_{u, n}
        }{
            \boldsymbol{\mathsf{B}}_{u, n}
        }
    \bigg].
\end{equation}
If the difference is positive (respectively negative), the evidence for the reduced (resp. original) model is greater, so it (resp. the posterior model) is selected as the updated model.

\section{Results and discussion}\label{sec:results}

Game transitions increase the non-stationarity of the environment beyond that caused by reciprocal adaptation, 
resulting in role reversals, strategic uncertainty, and eventual equilibrium selection.
We use a time-based linear interpolation of the payoff matrices of two games $\texttt{g}$ and $\texttt{g}'$ to transition between them. 
Given a time of transition $t_{\text{x}}$ and a transition duration $T_{\text{x}}$,
the preferences $p^*$ of each agent are updated each time step within the interval $(t_{\text{x}} - \frac{T_{\text{x}}}{2}) \le t \le (t_{\text{x}} + \frac{T_{\text{x}}}{2})$,
via mixing parameter 
$l = \big(t - (t_{\text{x}} - \frac{T_{\text{x}}}{2})\big)/T_{\text{x}}$,
with
\begin{equation}
p^* = \sigma \Big(
    (1 - l) \, \texttt{g} 
    \;+\; l \, \texttt{g}'
\Big).
\end{equation}

\subsection{VFE and strategic uncertainty}\label{sec:results-vfe}

\begin{figure}
    \centering
    \includegraphics[width=1\linewidth]{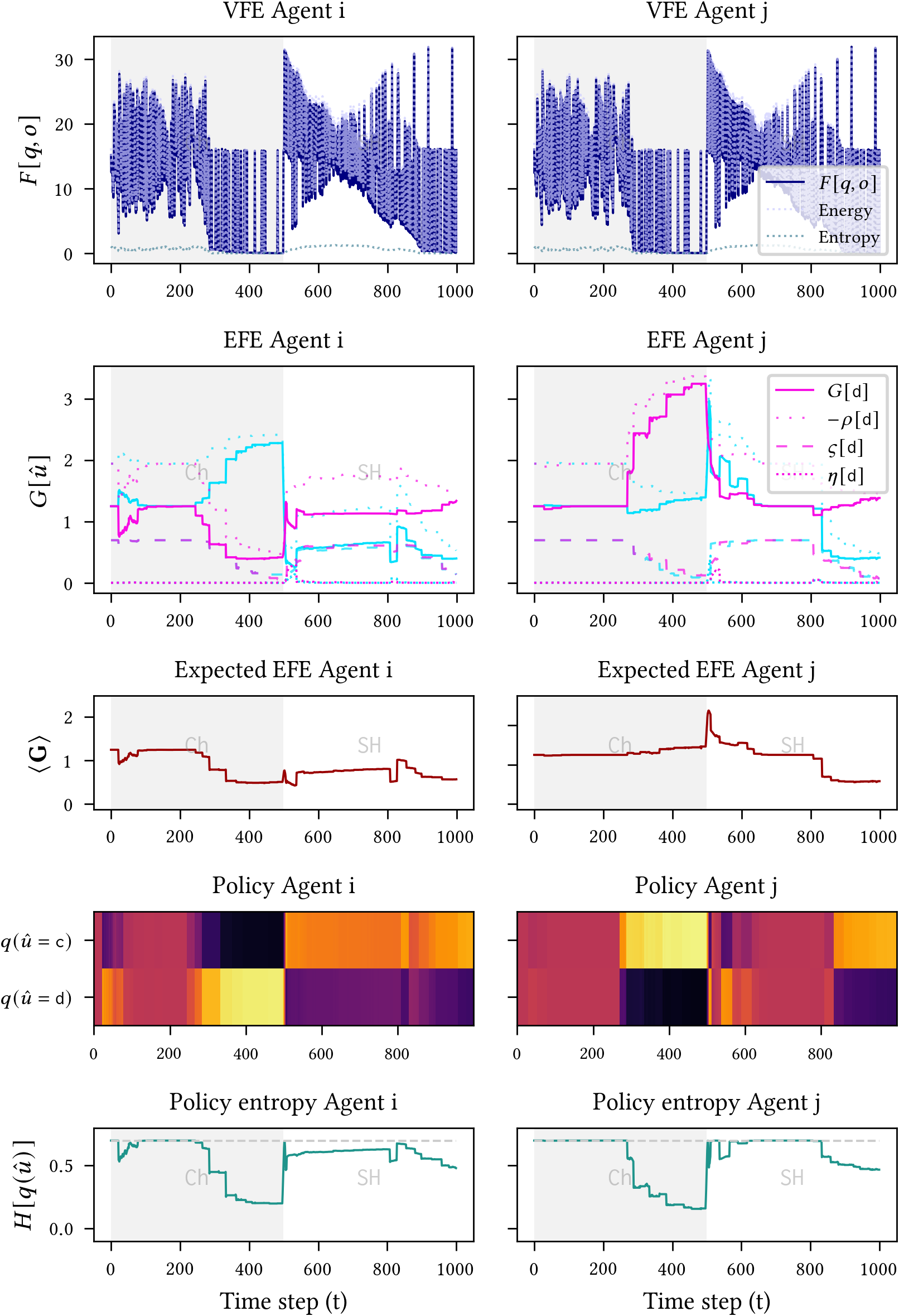}  
    \caption{
        Dynamics of a game transition with two agents ($\beta_1=15$). 
        500 steps of \texttt{Ch} followed by 500 steps of \texttt{SH} with a 10-step transition.
        In the EFE plots, blue represents `cooperate', and pink represents `defect'.
        In the policy heatmap plots, a lighter colour indicates a higher probability.
    }
    \label{fig:transition-timeseries}
\end{figure}

To illustrate the dynamics of transitioning between games, we run a two-player \texttt{Ch} game for 500 iterations, followed by a \texttt{SH} game for another 500, with payoffs as per Eq.~\ref{eq:2x2-payoffs}.
The transition occurs over $T_{\text{x}} = 10$ iterations.
The outcome is shown as time series plots for various quantities for each of the two agents in Figure~\ref{fig:transition-timeseries}.

\begin{figure}
    \centering
    \includegraphics[width=1\linewidth]{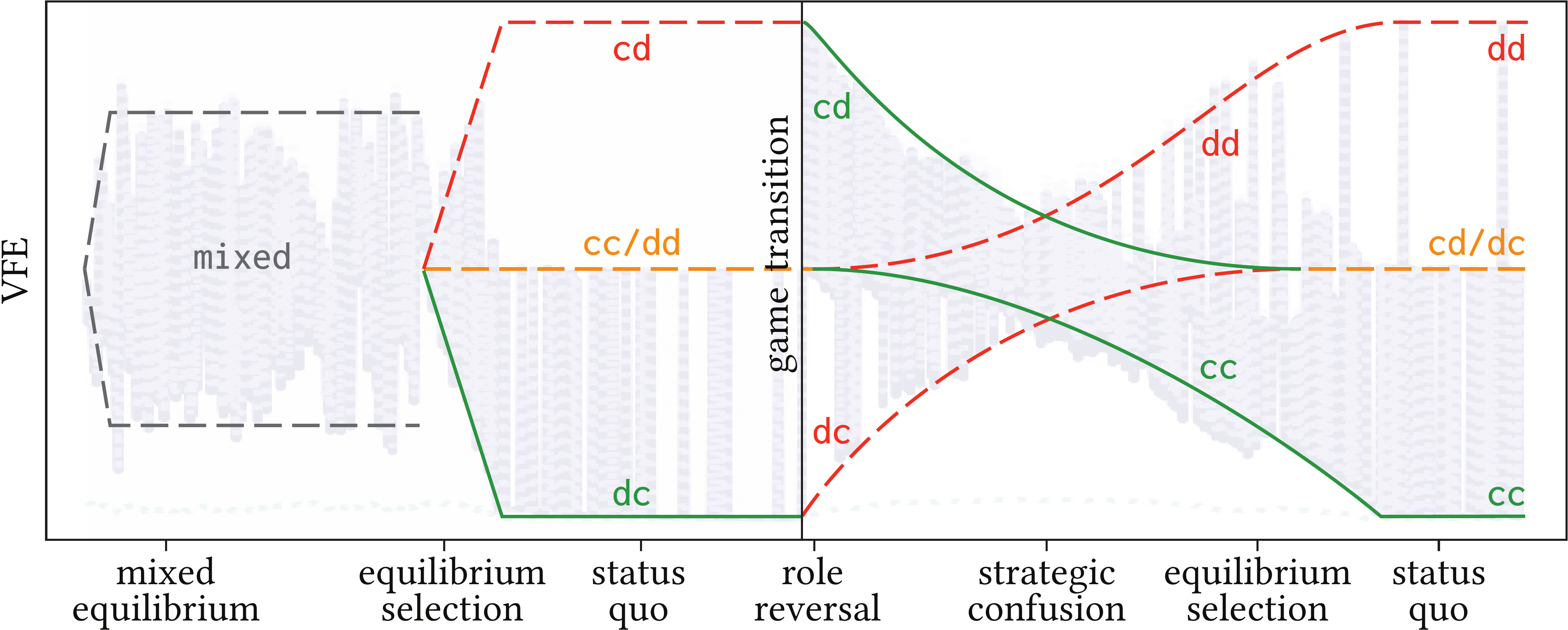}
    \caption{
        Stylised bounds on the dynamics of the VFE.
        }
    \label{fig:VFE-stylised}
\end{figure}

The VFE, $F[q, o]$, quantifies how surprising an observation $o$ is under the current beliefs $q(s)$.
As shown in Fig.~\ref{fig:transition-timeseries} (first row)---and further highlighted by the stylised bounds in Fig.~\ref{fig:VFE-stylised}---, the ensemble is initially in a mixed equilibrium. 
The symmetry is broken at $t\approx 250$, when, following a model parameter update, $i$'s policy moves towards defection, and $j$ appropriately responds by moving towards cooperation (incentivised by the game payoffs).
The ensemble remains in the selected $\texttt{dc}$ equilibrium for the remainder of the $\texttt{Ch}$ game.

If the agents' $\beta_1$ values were much higher, the VFE would follow the green stylised curve in Fig.~\ref{fig:VFE-stylised} much more closely.
However, the chosen $\beta_1 = 15$ value causes some `suboptimal' actions to be sampled from $q(\hat u)$ occasionally, so some observations deviate from the `status quo' (Fig.~\ref{fig:VFE-stylised}, green) in one (orange) or both (red) observation modalities, showing different levels of surprise (VFE) each\footnote{The VFE is additive over the factors of the generative model}.

After the game transition, there is a role reversal: the cooperating agent becomes a defector, and vice versa.
This is explained by the EFE, $G[\hat u]$ (Fig.~\ref{fig:transition-timeseries}, second row), where $i$'s (negative) pragmatic value of cooperating ($-\rho[\texttt{c}]$, blue) drops dramatically, surpassing the pragmatic value of defecting ($-\rho[\texttt{d}]$, pink).
This is because the preferences $p^*$ have now changed, and because $j$ has been cooperating up to that point---i.e., $i$ believes $j$ is a cooperator, that he can be trusted due to his reputation, making it more appealing for $i$ to cooperate as well.
However, the obverse is also at play for $j$, which transitions to defecting.
We note that this role reversal happens regardless of the transition duration, $T_{\text{x}}$.
With the transition, there is a sudden increase in the epistemic value (salience $\varsigma$ and novelty $\eta$) of actions, leading to exploratory behaviour.

As the agents persist with their new strategies, the pragmatic value of each action changes to reflect them.
The absolute difference $\big|\rho_j[\texttt{c}] - \rho_j[\texttt{d}]\big|$ diminishes, increasing the entropy of the policy, $q(\hat u_j)$, up to a point of maximal `strategic confusion', where the VFE is approximately the same for any possible observation (everything is just as surprising as anything else).
This confusion is resolved by equilibrium selection, with a small spike in novelty followed by a decrease in salience as the two agents' policies converge to their final values.

In this particular trial, the ensemble arrives at a payoff-dominant equilibrium with predominantly $\texttt{cc}$ strategies (modulated by the rationality of the agents).
But this may not always be the case, as we show next.

\subsection{EFE and equilibrium selection}\label{sec:results-efe}



The ensemble-level expected EFE, $\mathfrak{G} = \sum_i \langle \boldsymbol{\mathsf{G}} \rangle^{(i)}$, closely related to the joint objective recently proposed in \cite{Hyland2024FreeEnergyEquilibria}, provides a compact measure of the state of the ensemble over time.
By running several trials of an \ac{INFG} we can use this statistic to characterise the different equilibria of the game, as well as (an approximation of) the relative size of their basin of attraction.


\begin{figure}
    \centering
    \includegraphics[width=1\linewidth]{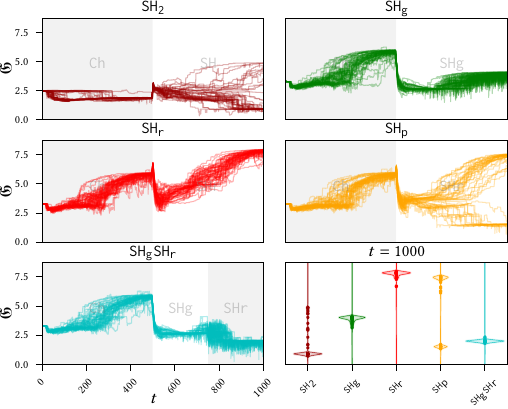}
    \caption{
        The ensemble-level expected EFE, $\mathfrak{G}$, highlights (the relative size of the basin of attraction of) the equilibria of a game ($\beta_1 = 30$).
        The bottom-right plot shows the kernel density estimate (Gaussian kernel, 0.08 bandwidth) of the PDF of final values under each condition.
        }
    \label{fig:EFE-equilibria}
\end{figure}


In Figure~\ref{fig:EFE-equilibria}, we show values of $\mathfrak{G}$ under different experimental conditions.
The data were obtained by running 50 trials of a chosen (sequence of) \ac{INFG} for each condition, which in the plots are shown superimposed (for a given condition).
Data points that are superimposed on the plots appear darker, highlighting the relative number of trials that take similar values.
The chosen \ac{INFG} are again \texttt{Ch} followed by \texttt{SH}, with a two-player condition ($\texttt{SH}_\texttt{2}$; as in Fig.~\ref{fig:transition-timeseries}), and the following three-player conditions using three variants of the $\texttt{SH}$ game:
\begin{itemize}
    \item a `green' variant ($\texttt{SH}_\texttt{g}$) where only two agents are required in order to successfully hunt a stag, 
    \item a `red' variant ($\texttt{SH}_\texttt{r}$) where all three agents are required, and
    \item a `penalty' variant ($\texttt{SH}_\texttt{p}$), where all three agents are required and the temptation to defect is lowered (by setting $T = P$).
\end{itemize}
Their respective payoff matrices are:
\begin{equation*}
\texttt{SH}_{\texttt{g}} =
    \begin{bmatrix}
        \begin{bmatrix}
          R & {\color{OliveGreen}R} \\ 
          {\color{OliveGreen}R} & S 
        \end{bmatrix},
        \begin{bmatrix}
          T & P \\ 
          P & P 
        \end{bmatrix}
    \end{bmatrix};
\;\;
\texttt{SH}_{\texttt{r}} =
    \begin{bmatrix}
        \begin{bmatrix}
          R & {\color{red}S} \\ 
          {\color{red}S} & S 
        \end{bmatrix},
        \begin{bmatrix}
          T & P \\ 
          P & P 
        \end{bmatrix}
    \end{bmatrix};
\end{equation*}
\begin{equation*}
\texttt{SH}_{\texttt{p}} =
    \begin{bmatrix}
        \begin{bmatrix}
          R & S \\ 
          S & S 
        \end{bmatrix},
        \begin{bmatrix}
          {\color{YellowOrange}P} & P \\ 
          P & P 
        \end{bmatrix}
    \end{bmatrix}
\end{equation*}
such that e.g. $\texttt{g}_{\texttt{SH}_{\texttt{g}}}(\texttt{c}, \texttt{d}, \texttt{c}) = R$, but  $\texttt{g}_{\texttt{SH}_{\texttt{r}}}(\texttt{c}, \texttt{d}, \texttt{c}) = S$,
with $R>T>P>S$ assigned the integers from 1 to 4.
The payoff matrix for the three-player $\texttt{Ch}$ has the same form as $\texttt{SH}_\texttt{r}$, except with $T>R>S>P$ (as per \S\ref{sec:infg}).
A fifth and final experimental condition is included, with an additional transition from $\texttt{SH}_\texttt{g}$ to $\texttt{SH}_\texttt{r}$ post-$\texttt{Ch}$. 

The five time-series plots (one for each condition) in Fig.~\ref{fig:EFE-equilibria} show superimposed series for 50 repeats. 
The bottom-right plot shows a side-by-side comparison of the values of $\mathfrak{G}$ at $t=1000$ under each condition. 

We first look at the $\texttt{Ch}$ period, where the ensemble starts at a mixed equilibrium ($t=0$) and reaches a pure equilibrium ($t \le 500$).
In the two-agent case, the ensemble tends toward one of $\texttt{cd}$ or $\texttt{dc}$ fairly quickly, although in one of the trials it stays in the mixed equilibrium through to the end.
There are no underlying configuration changes for $\texttt{Ch}$ between the three-agent cases, so any differences are caused by stochasticity.
In the final $\texttt{Ch}$ equilibrium, invariably, one of the agents defects and the rest cooperate (up to symmetry).
Since $\mathfrak{G}$ is additive, it decreases for the two-player condition, as only one of two agents has to choose the `worse' action, compared to the increase under the three-player conditions, where two-thirds of the ensemble select the `worse' action.
This is an instance where the Nash Equilibrium is not socially optimal.

The final $\texttt{Ch}$ equilibrium sets the prior conditions for the subsequent $\texttt{SH}$ game, amounting to a `pre-equilibrium'~\cite[p. 3]{Young2004InteractiveLearning}.
The $\texttt{SH}$ game has a \textit{risk-dominant equilibrium} (RDE) with a higher $\mathfrak{G}$ (i.e. worse overall), and a \textit{payoff-dominant equilibrium} (PDE) with a lower $\mathfrak{G}$ (i.e. better overall).
In the $\texttt{SH}_\texttt{2}$ and $\texttt{SH}_\texttt{p}$ conditions, we see a bifurcation occur where a portion of the trials ends in each equilibrium, with the majority of $\texttt{SH}_\texttt{2}$ (resp. $\texttt{SH}_\texttt{p}$) ending in the PDE (resp. RDE).
This shows the relative size of their basins of attraction (noting that $\texttt{SH}_\texttt{2}$ may need beyond $t = 1000$ to converge further).

All the $\texttt{SH}_\texttt{g}$ trials end in the PDE; all of $\texttt{SH}_\texttt{r}$, in the RDE.
This is somewhat paradoxical: when two are required to cooperate ($\texttt{SH}_\texttt{g}$), all three cooperate; conversely, when three are required to cooperate ($\texttt{SH}_\texttt{r}$), none cooperates.
Interestingly, the $\texttt{SH}_\texttt{p}$ variant could be seen as an attempt to direct the ensemble towards a better equilibrium via penalising certain behaviour (cf. mechanism design), with mild success.
On the other hand, including an interim transition through $\texttt{SH}_\texttt{g}$ generates trust and thus `bootstraps' cooperative behaviour, stewarding the collective~\cite{Bak-Coleman2021StewardshipGlobal} towards the PDE without needing to resort to penalties.

\section{Conclusion}

We have proposed a factorisation of the generative model of \ac{AIF} agents that brings the framework in closer alignment with game theory, particularly for multi-agent interactions.
In this factorisation, each agent has explicit, individual-level beliefs about the internal states of others, and uses them to plan strategically in a joint (game-theoretic) context.
This allows ego to flexibly track the internal states of others outside the joint interaction context, and incorporate that information as required by the interaction.
This would be particularly useful when the agents involved in a given interaction change or if the agent participates in multiple interactions at a given time (cf. network games).

We have applied our proposed model to two- and three-agent \ac{INFG} and included transitions between games that the agents have to adapt to.
We have shown how the VFE and EFE can be used to analyse the dynamics of ensembles of agents interacting strategically~\cite{Barfuss2022DynamicalSystems}---in particular, to highlight the equilibria of games and the relative size of their attractors---, and provided an example where this is used to motivate an intervention leading the ensemble to a better outcome based on trust, rather than punishment.
The use of these measures may help in the conceptualization of groups of agents as collective agents with self-organizing dynamics and operational closure~\cite{Ramstead2021MultiscaleIntegration, Kiverstein2022ProblemMeaning, Hyland2024FreeEnergyEquilibria}.

\ac{AIF} and game theory applied to multi-agent systems offer a rich theoretical and experimental landscape for exploring adaptive behaviours in intelligent agent interactions. 
This intersection of cognitive science and artificial intelligence not only provides insights into individual decision-making but also paves the way for understanding the collective dynamics that shape social behaviour in complex environments.

Ego's beliefs about her own policy $q(\hat u_i)$ reflect a form of introspection: inferring one's internal mental states by observing one's actions~\cite{Harre2022WhatCan, Sandved-Smith2021ComputationalPhenomenology}. This is contrasted with interoception: the perception of internal bodily sensations, which would provide direct access to internal states like $q(\hat u_i)$. By endowing ego with interoceptive access to $q(\hat u_i)$, one could bypass the need for further inference about her internal state in future steps of the model.
Future work shall explore how learning the observation model could capture the rationality of an opponent; the potential for more complex transition models conditioned on the actions of all agents, or for modelling other hidden variables such as opponent preferences~\cite{Ruiz-Serra2023InverseReinforcement, Chan2021ScalableBayesian}; or the effects of different EFE formulations~\cite{Hafner2022ActionPerception, Millidge2021WhenceExpected, Champion2024ReframingExpected} on game outcomes.

\begin{acks}
JRS is supported by an Australian Government Research Training Program (RTP) Scholarship.
We acknowledge the Gadi people of the Eora nation as the traditional custodians of the land on which The University of Sydney now stands, and that their sovereignty was never ceded.
\end{acks}



\bibliographystyle{ACM-Reference-Format} 
\balance
\bibliography{references}


\end{document}